# Is the Kinematics of Special Relativity incomplete?

Ernst Karl Kunst

**An analysis of composite inertial motion (relativistic sum) within the framework of special relativity leads to the conclusion that every translational motion must be the symmetrically composite relativistic sum of a finite number of quanta of velocity. It is shown that the resulting space-time geometry is Gaussian and the four-vector calculus has its roots in the complex-number algebra, furthermore, that Einstein's "relativity of simultaneity" is based on a misinterpretation of the principle of relativity. Among others predictions of the experimentally verified rise of the interaction-radii of hadrons in high energetic collisions are derived. From the theory also follows the equivalence of relativistically dilated time and relativistic mass as well as the existence of a quantum of time (fundamental length) and its quantitative value, to be found in good accord with experiment**

## Introduction

Modern physics developed experimental methods the results of which in principle confirm special relativity as proposed by Albert Einstein as well as its further mathematical shaping mainly by Hermann Minkowski. At the same time new physical phenomena were discovered in high energy (collider) physics which usually are not brought in connection with the kinematics and mechanics of special relativity. These phenomena, for which a convincing physical explanation has not been found yet, can be grouped as follows:

1) The rise of the interaction-radius and total cross-section of elementary particles (hadrons) with increasing energy (of the beam);

2) The shrinkage of mean-free-paths of ultra relativistic particles (nucleii) in material media;

3) The obvious existence of shortest life-times of particle resonances of the order $10^{-24}$ s,

where "s" means second. As widely known, special relativity rests on two premises:

The invariance of the physical laws for all observers, independent of the state of inertial motion (principle of relativity);

The constancy of the velocity of light in a frame of rest independent of the velocity of the source,

wherefrom the Lorentz transformation results automatically.

A reconsideration of the kinematics of special relativity results in a novel definition of the concept of velocity between any two inertial frames of reference and a modification of the Lorentz transformation. In some aspects the predictions of this new kinematic view deviate markedly from special relativity especially at velocities near that of light and, together with a review of the "relativity of simultaneity", explain the previously mentioned experimentally verified physical phenomena both qualitatively and quantitatively as being of relativistic origin.

## 1. Is Inertial Motion Quantized?

Imagine a system $S_2$ ($x_2$, $y_2$, $z_2$, $t_2$) moving inertially at constant uniform speed "w" parallel to a system $\Sigma$ (x, y, z, t) and the latter moving at the velocity "v" relative to an observer resting in the coordinate source of a sytem $S_1$ ($x_1$, $y_1$, $z_1$, $t_1$) "at rest", according to the principle of relativity.

1) It is demonstrated that the resulting relativistically composite velocity $u = (v + w)/(1 + vw/c^2)$ of $S_2$ - as observed from $S_1$ - is variable, dependent on the respective value of v and w, where but $|v| + |w|$ is always constant. Einstein considered

$$u = c \frac{2c - \lambda - \kappa}{2c - \lambda - \kappa + \frac{\lambda \kappa}{c}} < c \;, \tag{1}$$

in order to prove that the relativistic sum of two velocities which are slower than light always results in a velocity slower than light [1]. We posit $v = c - \lambda$, $w = c - \kappa$ and $\lambda_{max} = \lambda + \kappa$, whereby always $\lambda_{max} > (\lambda_{max} - \lambda) > 0$. From (1) follows

$$u = c \frac{2c - \lambda_{max}}{2c - \lambda_{max} + \frac{(\lambda_{max} - \lambda)\lambda}{c}} < c = \text{Extrem.}$$

if the postulates are satisfied $0 < (\lambda_{max} = \text{const}) \leq 2c$, $\lambda \leq c$, $\kappa \leq c$. Clearly u reaches a maximum value if $(\lambda_{max} - \lambda)\lambda/c = \text{Min}$. This is the case if $(\lambda_{max} - \lambda)$, or $\lambda$, reaches its maximum value. The composite velocity u reaches its minimum value if $\lambda = \kappa$ and $\lambda_{max} = 2\lambda = 2\kappa$:

$$u = c\frac{2c - \lambda_{max}}{2c - \lambda_{max} + \frac{\lambda^2_{max}}{4c}} = \frac{2c(c-\lambda)}{2(c-\lambda) + \frac{\lambda^2}{c}} < c = \text{Min.},$$

which attains the form

$$u = \frac{2v_1}{1 + \frac{v_1^2}{c^2}} = \text{Min.} \tag{2}$$

if $c - \lambda = v_1$.

2) If vice versa $u = $ const it seems clear that any points $\Sigma$ could simultaneously exist (can be thought of or physically realized) between $S_1$ and $S_2$ so that any relativistic sum $(v + w)/(1 + vw/c^2) = $ const conceivably would yield u with $|v| + |w| = $ Max. $= 2v_1 > u$ if (2) is valid and $|v| + |w| = $ Min. $= u < 2v_1$ if $\Sigma$ coincides with $S_1$ or $S_2$. Thus - considering the extremes only -, apparently is valid:
$\forall u\ (u = (v + w)/(1 + vw/c^2) = \text{const} \wedge |v| + |w| = u \wedge |v| + |w| = 2v_1)$.

3) Now we maintain that the symmetrically composite velocity (2) be the only existing proper velocity of $S_2$ relative to $S_1$ implying both systems to move symmetrically at equal but oppositely directed velocity $v_1$ relative to a point in space-time considered to be at rest, now designated $\Sigma_0$:
$\exists u\ (u = 2v_1/(1 + v_1^2/c^2) \wedge |v| + |w| = \text{Max.} = 2v_1)$.

4) Consider 2) to be true. In this case the velocity of $\Sigma_0$ relative to $S_1$ will be $v_1$ and the distance $\bar{S_1}\bar{S_2}$, as observed from $\Sigma_0$, for reasons of symmetry $2v_1 \times \Delta t_0$. Thus, it must be valid

$$u \times \Delta t_1 = 2v_1 \times \Delta t_0, \tag{2a}$$

whereby u in the left-hand member according to 2) apparently could be non-composite and simultaneously composite. If a light signal is transmitted from $S_2$ to $S_1$ via $\Sigma_0$ it shall travel the distance between those systems in the time

$$\frac{u \times \Delta t_1}{c} = \frac{2v_1 \times \Delta t_0}{c}. \tag{2b}$$

The light signal must in any case propagate via $\Sigma_0$. A non-composite u does not exist in the space-time of $\Sigma_0$, respectively contain $\Sigma_0$. This requires u in the left-hand member to be symmetrically composite. Thus, u to be non-composite is ruled

out - as well as any other composite velocity with no point $\Sigma_0$. Therefore, 2) must be false and 3) true so that (2b) attains the only possible form

$$\frac{2v_1 \times \Delta t_1}{c\left(1 + \frac{v_1^2}{c^2}\right)} = \frac{2v_1 \times \Delta t_0}{c}, \qquad (2c)$$

which guarantees the light signal to propagate via $\Sigma_0$.

5) Hence between any two inertial frames of reference $S_1$ and $S_2$, moving relative to each other with constant uniform speed, an inherent preferred reference point $\Sigma_0$ always exists at rest  - for the time of the translational motion - relative to $S_1$ and $S_2$ implying their velocity relative to $\Sigma_0$ to be symmetrically equal and oppositely directed and relative to another permanently a relativistic sum (2).

The chain of argumentation and the result that any velocity u be symmetrically composite also is valid for the velocities "$v_1$", "$v_2$" etc. Thus, a point $\Sigma_1$ also must exist between $\Sigma_0$ and $S_2$, a point $\Sigma_2$ between $\Sigma_0$ and $\Sigma_1$, $\Sigma_1$ and $S_1$, or $S_2$ and so forth so that it is clear that

$$v_0 = \frac{2v_1}{1 + \beta_1^2} , \ v_1 = \frac{2v_2}{1 + \beta_2^2} ,..., v_{n-1} = \frac{2v_n}{1 + \beta_n^2} , \qquad (3)$$

where $\beta_1 = v_1/c$, $\beta_2 = v_2/c$,..., $\beta_n = v_n/c$. Hence the relative translational movement between any two inertial systems $S_1$ and $S_2$ must be the relativistic sum of $2^n$ quanta of velocity. Because

$$v_{n-1} = \frac{2v_n}{1 + \beta_n^2}, \ v_{n-2} = \frac{2^2 v_n}{(1 + \beta_n^2) \times (1 + \beta_{n-1}^2)}, \ v_{n-3} = \frac{2^3 v_n}{(1 + \beta_n^2) \times (1 + \beta_{n-1}^2) \times (1 + \beta_{n-2}^2)} \text{ etc.}$$

the product results

$$v_0 = \frac{2^i v_i}{\prod_{n=1}^{i}(1 + \beta_i^2)} = \frac{2^i v_i}{N_i} , \qquad (4)$$

where i = 1, 2,..., n; $v_0$ designates henceforth naturally composite velocity and it is clear that this form of velocity quantization maintains the group property of space-time. It follows that $2^n = N_n$ implies $v_0 = c$.
The fact that every velocity is quantized in the order of (3) and all elements of (3)

without any exception have a common attribute $v_k = 2v_{k+1}/(1 + \beta^2_{k+1})$, where k = 0, 1..., n - 1, implies the laws of well-arranged sets to be applicable or with other words: (3) is a well-arranged set. The well-known definition of well-arranged sets applied on (3) implies that a first minimum composite velocity "$v_{min}$" different from null exists. It follows that (4) is finite and $v_{min} \times 1s = c \times \tau_0$ or

$$\tau_0 = \frac{v_{min}}{c}, \qquad (5)$$

where "$\tau_0$" means quantum of time. The further development of the theory allows the verification of the existence and exact derivation of the quantitative value of $\tau_0$. Thus, velocity is always composite according to (4), which usually falsely is taken to be non-composite. It is clear that even in the ultra relativistic region, where $v_0 \to c$, direct measurements would unveil no difference to the classical apparently non composite velocity "v". But this cannot be true for measurements of momentum or energy, which are based on the electron Volt (eV). A protron (antiprotron), accelerated in an electrical field with the potential difference of one eV, would reach the subrelativistic velocity $v_0 \approx v \approx 3 \times 10^4$ cm/s. Thus, with rising energy or momentum a systematic deviation of the correct composite value according to (4) from the special relativistic one of the order of magnitude

$$\frac{|n \times eV|}{c} = p = mv\gamma = p_0 \times N_i = mv_0\gamma_0 \times N_i \qquad (6)$$

must be taken into consideration, where "n" means any number (multiplicity) of eV, $p_0$ momentum on the strength of naturally composite velocity according to (4), p special relativistic momentum, m rest mass, v conventional velocity, $\gamma$ and $\gamma_0$ Lorentz factor on the strength of the conventional and the composite concept of velocity, respectively.

According to (4) must be valid $v_0 \to v_1, v_2, v_3,..., v_n$ if $v_0 \to c$ so that $N_i$ successively reaches the values

$$\begin{aligned}
N_1 &= 1 + \beta_1^2 \to 2, \text{ when } \beta_1 \to 1, \\
N_2 &= 1 + \beta_2^2 + 4\beta_2^2(1 + \beta_2^2)^{-1} \to 4, \text{ when } \beta_2 \to 1, \\
N_3 &= 1 + \beta_3^2 + 4\beta_3^2(1 + \beta_3^2)^{-1} + 16\beta_3^2(1 + \beta_3^2 + 4\beta_3^2(1 + \beta_3^2)^{-1})^{-1} \to 8, \text{ when } \beta_3 \to 1 \text{ etc.}
\end{aligned} \qquad (6a)$$

This means that $N_i$ always can be approximated as is shown below. Given the foregoing it is clear that relative to the preferred frame of reference $\Sigma_0$, being the kinematic center, the space-time of $S_1$ and $S_2$ must be strictly symmetrically equal.

2. <u>The Symmetric Lorentz Transformation</u>

Consider the inertial systems $S_1$ and $S_2$ moving uniformly parallel and oppositely directed relative to the natural frame $\Sigma_0$ at rest relative to them. The symmetric transformation equations are derived by assuming the validity of:

   1) The Lorentz transformation (principle of relativity),

   2) The inherent rest frame of nature $\Sigma_0$ at rest in any translational movement implying the absolute equality of the inertial systems under consideration  (principle of symmetry).

It is understood that the bodies resting in the coordinate sources of $S_1$ and $S_2$ are geometrically identical if they are compared with each other at rest, according to the Einsteinian definition [2]. According to postulate 2) must the transformation be absolutely symmetric in respect to the systems under consideration. Furthermore, according to both postulates observers resting in $S_1$ and $S_2$ must consider themselves at rest and at the same time to move relative to $\Sigma_0$ and the other system. Thus, the observer in $S_1$ will besides the Lorentz transformation according to postulate 1) - first line of (7) -, where he considers himself at rest, deduce a second transformation from the moving frame $S_2$ - according to the principle of relativity now considered at rest - back to his own system - now considered moving relative to $\Sigma_0$ and $S_2$ (see Fig. 1):

$$x_2' = \gamma_0(x_1 - v_0 t_1), \quad y_2' = y_1, \quad z_2' = z_1, \quad t_2' = \gamma_0(t_1 - \frac{v_0}{c^2}x_1),$$

$$x_1^\circ = \gamma_0(x_2' + v_0 t_2'), \quad y_1^\circ = y_2', \quad z_1^\circ = z_2', \quad t_1^\circ = \gamma_0(t_2' + \frac{v_0}{c^2}x_2'),$$

(7)

whereby

$$\gamma_0 = \frac{1}{\sqrt{1-\beta_0^2}} \;.$$

The dashes designate the moving system $S_2$ and the open circles the reference rest frame $S_1$, now considered moving. Likewise the observer resting in $S_2$ will deduce:

$$x_1' = \gamma_0(x_2 + v_0 t_2), \quad y_1' = y_2, \quad z_1' = z_2, \quad t_1' = \gamma_0(t_2 + \frac{v_0}{c^2}x_2),$$

$$x_2^\circ = \gamma_0(x_1' - v_0 t_1'), \quad y_2^\circ = y_1', \quad z_2^\circ = z_1', \quad t_2^\circ = \gamma_0(t_1' - \frac{v_0}{c^2}x_1').$$

(8)

According to presupposition 2) for the observer resting in either system is valid:

$$x_2 = x_1, \quad y_2 = y_1, \quad z_2 = z_1, \quad t_2 = t_1,$$
$$x_1' = x_2', \quad y_1' = y_2', \quad z_1' = z_2', \quad t_1' = t_2', \quad (9)$$
$$\overset{\circ}{x_2} = \overset{\circ}{x_1}, \quad \overset{\circ}{y_2} = \overset{\circ}{y_1}, \quad \overset{\circ}{z_2} = \overset{\circ}{z_1}, \quad \overset{\circ}{t_2} = \overset{\circ}{t_1}$$

and always $|v_0| = |-v_0|$. If the upper lines of (7) and (8) are inserted into the second lines, the identity results:

$$\overset{\circ}{x_1} \equiv x_1, \quad \overset{\circ}{t_1} \equiv t_1,$$
$$\overset{\circ}{x_2} \equiv x_2, \quad \overset{\circ}{t_2} \equiv t_2. \quad (10)$$

Equs. (7) to (9) have been deduced by strictly considering transformations from a system considered to be at rest to the one considered to move. The different states of motion have intentionally been made distinguishable by the use of different symbols. The proper inverse Lorentz transformation in (7) and (8) is given by the respective second line, where the former moving system now must be considered to be at rest according to the principle of relativity. The invariance of the scalar

$$x_2'^2 + y_2'^2 + z_2'^2 - c^2 t_2'^2 = x_1^2 + y_1^2 + z_1^2 - c^2 t_1^2 \quad (11)$$

follows from (7) to (10).

### 3. The Symmetric Minkowski-Diagram; Equivalence of Four-Vector and Complex Number Calculus

Consider the preferred frame of reference of nature $\Sigma_o$, relative to which $S_1$ and $S_2$ are moving at oppositely directed velocity $v_1$, and the inertial frame $S_2'$ propagating relative to $S_1$ at the velocity $v_0$, and their paths in space-time (world-lines). Fig. 1 shows a diagram of space-time on the grounds of (7) to (10). It is evident that due to the absolute symmetry of $S_1$ ($S_1'$) and $S_2$ ($S_2'$) relative to $\Sigma_0$ the triangles $(0, S_1, S_2')$ and $(0, S_2, S_1')$ must be Pythagorean ones.

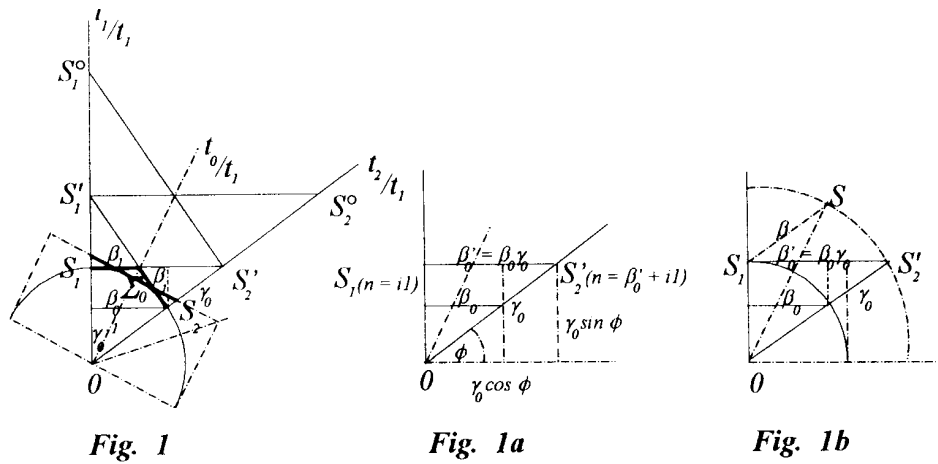

Fig. 1     Fig. 1a     Fig. 1b

From Fig. 1 results directly the Pythagorean relation

$$\beta_1^2 - (\beta_0 - \beta_1)^2 - (1 - \gamma_0^{-1})^2 = 0.$$

Computation yields $v_0 = 2v_1/(1 + \beta_1^2)$ and, therewith, the geometrical proof of (2), whereby $\overline{0S_1} = \overline{0\Sigma_0} = 1$ and $u = v_0$. Furthermore, from the diagram

$$\frac{\overline{0S_2'}}{\overline{0S_1}} = \frac{\overline{0S_2'}}{\overline{0\Sigma_0}} = \gamma_0$$

follows, implying between $\Sigma_0$ and $S_1'$, or $S_2'$, the Minkowskian space-time relations and, thus, the special relativistic transformation

$$ct' = ct_0 \cosh \alpha - x_0 \sinh \alpha,$$
$$x' = x_0 \cosh \alpha - ct_0 \sinh \alpha$$

to be valid, where $\cosh \alpha = \gamma$, $\sinh \alpha = \beta\gamma$, $\beta = v/c$.

As Fig. 1a shows is the point $S_2'$ of the Pythagorean triangle $(0, S_1, S_2')$ the complex number $n(\beta_0', 1)$ in a complex $ct, x$-plane of space-time - we abstract from $y$ and $z$ -, written $n = \beta_0' + i1$ or in polar notation

$$n = \gamma_0 (\cos\Phi + i \sin\Phi).$$

Because $|n| = (\gamma_0^2(\cos^2 \varphi + \sin^2 \varphi))^{1/2}$ and $\cos^2 \varphi + \sin^2 \varphi = 1$, it follows

$$|n| = \sqrt{\gamma_0^2} = \gamma_0,$$

the minimum being $|n| = 1$. If dashed symbols denote physical quantities in $S_2'$ and the undashed ones the respective values in $S_1$, considered to be at rest, it must be valid

$$|n| = \gamma_0 = \frac{\beta_0'}{\beta_0} = \frac{E'}{E} = \frac{m'}{m} = \frac{dt'}{dt}, \qquad (12)$$

whereby $\beta_0'$ results from $v_0 dt'/(cdt)$.

As Fig. 1b shows will the special-relativistic space-like world-line $S_1^- S$ through multiplication by the Lorentz-factor γ be stretched to the (symmetric) world-line $S_1^- S_2^+$ - apart from a factor $N_i^{-1}$ according to (4) - and, therewith, the Minkowskian triangle (0, S, $S_1$) be transformed into the Pythagorean one (0, $S_2'$, $S_1$). In special relativity this is achieved by introduction of

$$\frac{dx}{ds}, \frac{dy}{ds}, \frac{dz}{ds}, \frac{dtc}{ds}, \tag{13}$$

whereby

$$-\frac{dx^2}{ds^2} - \frac{dy^2}{ds^2} - \frac{dz^2}{ds^2} + \frac{dt^2 c^2}{ds^2} = 1$$

and

$$\frac{ds}{dt} = (1-\beta^2)^{\frac{1}{2}} = \gamma^{-1}$$

so that the four components (13) attain the form

$$v_x \gamma, v_y \gamma, v_z \gamma, c\gamma$$

of four-velocity. Thus, the ct, x-plane of the space-time of special relativity is transformed into a Gaussian plane and the complex number algebra made applicable in the form of the four-vector calculus. Especially result from the complex number algebra directly the inner product of two four vectors q × r = q' × r', the square of two four-vectors $(q + r)^2 = (|q| + |r|)^2$, furthermore, the relativistic conservation of energy and momentum

$$E'^2 - E'^2 \beta_0^2 - E^2 = m'^2 c^4 - m'^2 c^2 v_0^2 - m^2 c^4 = 0 \tag{14}$$

or the well-known invariant

$$E = c \left( \frac{E'^2}{c^2} - p_0^2 \right)^{\frac{1}{2}} \tag{14a}$$

in the usual three-dimensional notation etc., where $p_0$ means momentum.
If E' and E'$\beta_0$ (→ E') coincide, evidently E = 0 or m = 0 follows, the rest energy or mass of the photon. It is clear that, owing to the absolute symmetry relative to $\Sigma_0$, E' in its lowest approximation ($N_i \approx 1$) means twice the center-of-mass energy if material bodies of identical mass are considered and $v_0 \ll c$. This also is the "natural" order of magnitude of the photon energy as Fig. 1 shows.

### 4. Is Einstein's "Relativity of Simultaneity" Correct ?

The empirical principle of relativity basically implies that any observer in whatever state of inertial motion relative to another system has to consider himself at rest in his frame of rest, with the consequence (among others) that light travels isotropically in all systems of reference alike, which fact is expressed by (11). Thus, it is clear that the hypothesis of FitzGerald and Lorentz that moving bodies are contracted by the factor $\gamma^{-1}$ in the direction of motion is not needed in special relativity to account for the null-result of the Michelson-Morley experiment on moving Earth. The principle of relativity rather explains solely and completely the outcome of this and similiar experiments.

1) Nevertheless, as widely known, introduced Einstein in special relativity the "relativity of simultaneity" to receive a "measurement rule", which allows for the "FitzGerald-Lorentz contraction" as a result of the theory. For this he maintained [2]:

i) that the length l' of a "moving" rod $r_{AB}$, as measured in the "moving" system, equals "according to the principle of relativity" the length l of a like rod (as compared at rest), resting relative to the former one in a system "at rest". Thus, the "moving" observer ($r_{AB}$) would find clocks (A and B), which are at rest relative to and synchronized in the system "at rest", not to be synchronous

ii) so that must be valid

$$t_B - t_A = \frac{r_{AB}}{c - v} \text{ and } t_{A'} - t_B = \frac{r_{AB}}{c + v}.$$

The difference $t_B - t_A$ means the time a light signal emitted in the system "at rest" needs from A to B, to be there reflected and travelling back to and reach A at the time $t_{A'}$, where $t_B - t_A = t_{A'} - t_B$.

2) It has been overlooked ever since that this proceeding is not admissible in the framework of special relativity and violates the very basis of that theory: the principle of relativity.
If the observer changes from a "resting" system to a "moving" one, then the latter becomes according to the principle of relativity the observers reference rest frame at rest. This implies that the former resting system must now be considered moving relative to the new rest frame, as also follows from (4), where $v_0$ results from successive Lorentz transformations in respect to the observer at rest. Thus, the time intervals $t_B - t_A = t_{A'} - t_B$ now belong to a moving system. If the rod resting in the system now considered at rest is designated $r_{ab}$ (to distinguish it from the now moving rod $r_{AB}$), it must be valid $r_{ab}/c = t_b - t_a = t_{a'} - t_b$, as measured by an observer resting at $r_{ab}$.

3) A correct transformation of the space-time coordinates of a light signal moving relative to the coordinate source of any inertial system into the respective coordinates of another system, being the rest frame of the observer, requires in any case the application of the addition theorem. This implies that conclusion ii) of 1) must be wrong. Instead the only possible and correct comparison of the lenghts of the rods $r_{AB}$ and $r_{ab}$ by the observer resting at $r_{ab}$ can be made if - at the time when the points "A" and "a" briefly coincide and the light signal is emitted in the system $r_{AB}$ according to ii) - the travel times are compared. Because $\pm c$ = const relative to $r_{AB}$ and $(v \pm c)/(1 \pm vc/c^2) = \pm c$ = const relative to $r_{ab}$, ii) correctly becomes

$$t_B - t_A = \gamma(t_b - t_a) = \frac{\gamma r_{ab}}{c} \text{ and } t_{A'} - t_B = \gamma(t_{a'} - t_b) = \frac{\gamma r_{ab}}{c}.$$

4) This result also can be demonstrated in a more precise way by a thought experiment, based on the reversal of Einstein's definition of simultaneousness: "Two events at different places of a frame of reference (considered) at rest are simultaneous if they can be seen simultaneously in the center of the connection line [3]."
Consider two systems in parallel inertial motion at velocity $v_0$, the length of a rigid rod resting in the coordinate source of the moving one being L' and L of another one of equal condition (as compared at rest), resting in the system considered at rest. When they pass each other and the marked centers of both rods briefly coincide at $t_{L'} = t_L = 0$, a light signal shall be transmitted oppositely directed in the system L' and trigger simultaneously e. g. a laser signal from either end of the rod in the direction of the rod L. According to the principle of relativity the light travels isotropically with the velocity c = const in the x-direction and -c = const in the -x-direction of L' and needs the time $L'/(2c) = \Delta t'/2$ to either end of the rod. Relative to the marked center of the resting rod L those light signals travel at velocity $(v_0 + c)/(1 + v_0c/c^2) = c$ parallel and $(v_0 - c)/(1 - v_0c/c^2) = -c$ antiparallel to the vector of velocity (coinciding with the x-direction), consequently isotropically, too, to reach either end of L after the time $L/(2c) = \Delta t/2$. Therefore, owing to the fact that $\Delta t' = \Delta t \gamma_0$, as observed from the system L at rest, an absolute symmetry of the propagation of the light signals from the respective center is found:

$$\frac{L'}{2c} + \frac{L'}{2c} = \frac{\Delta t \gamma_0}{2} + \frac{\Delta t \gamma_0}{2} \text{ and } \frac{L}{2c} + \frac{L}{2c} = \frac{\Delta t}{2} + \frac{\Delta t}{2}.$$

As observed from the resting system, the oppositely directed signals in the moving system also arrive simultaneously at both ends to trigger the laser signal, though dilated by the factor $\Delta t(\gamma_0 - 1)/2$, as compared with the simultaneous arrival of the light at either end of the resting rod after the time $\Delta t/2$. Multiplication by c delivers

$$\frac{L'}{2} + \frac{L'}{2} = \frac{c\Delta t \gamma_0}{2} + \frac{c\Delta t \gamma_0}{2} = \frac{L\gamma_0}{2} + \frac{L\gamma_0}{2}.$$

Thus, it follows conclusively that the length L' of the moving rod must be expanded by the factor $\gamma_0$ so that $L' = L\gamma_0$, as observed from the system considered at rest. But it already is from the symmetric setup of (7) to (10) clear that any interpretation of the latter transformation equations of the x- coordinate other than to take them at face value - which results in an expansion of the x-dimension - is ruled out.

5) The reasoning 2) - 4) also is in full accord with the well-known experiment of Fizeau in 1851, which in principle corresponds to Einstein's thought experiment 1), with the deviation that the velocity of light c' in the moving system (running water) is slower than in vacuo: c' = c/n, where "n" means refractive index. It has been overlooked by Einstein and ever since that Fizeau's result directly contradicts 1), because it is readily explained by the relativistic addition theorem [4] - in accord with 2) - 4). From the latter directly follows Fizeau's empirical formula and, therewith, the velocity of light in the moving medium to be c' = c/n - in vacuo n = 1 and c' = c -, whereas Einstein's argumentation 1) and the derived relations ii) would lead to the pre-relativistic result c' ≠ c/n.

Hence for an observer resting in the coordinate source of $S_1$ evidently the spatial dimensions of a body resting in the the moving system $S_2'$ are given by

$$\Delta x_2' = \gamma_0 \Delta x_1, \quad \Delta y_2' = \Delta y_1, \quad \Delta z_2' = \Delta z_1. \tag{15}$$

Thus, results

$$V_2' = \gamma_0 V_1, \tag{16}$$

where V means volume. Of course, an observer who happens to rest in the system $S_2$ will deduce the corresponding result:

$$V_1' = \gamma_0 V_2. \tag{17}$$

Hence (12) also is valid for the volume of a moving body: $|n| = \gamma_0 = V'/V$. Now let a system $S_2°$, propagating "within" a "real resting" system $S_1$, come to a halt within and relative to $S_1$, e. g. a material particle within some solid material. From (11) in connection with (12) we receive

$$x_2° = \gamma_0^2(x_1 - 2v_0 t_1 - \frac{v_0}{c^2}x_1), \quad y_2° = y_1, \quad z_2° = z_1,$$

wherefrom is deduced

$$\Delta x_2° = \gamma_0^2 \Delta x_1, \quad \Delta y_2° = \Delta y_1, \quad \Delta z_2° = \Delta z_1,$$
$$V_2° = \gamma_0^2 V_1 \tag{18}$$

and in the inverse case

$$V_1^\circ = \gamma_0^2 V_2 . \tag{19}$$

All other known special relativistic (optical and electrodynamical) effects also result from the first lines of (7) and (8) - owing to the coincidence with the Einsteinian Lorentz transformation.

### 5. Interaction-Radii and Geometrical Cross-Sections of Material Bodies in Ultra Relativistic Collision Events

If this theory is correct the relativistic expansion of length or volume should be noticable in ultra relativistic collisions of material particles. Because nearly all collision events in high energy physics more or less are of a grazing kind the mean geometrical dimensions of the colliding particles must be averages over all three axes x, y, and z of the according to (16) and (17) relativistically enlarged volumina. Especially the mean of the x-dimension must be

$$\overline{\Delta x_2'} = V_2'^{\frac{1}{3}} = V_1^{\frac{1}{3}} \gamma_0^{\frac{1}{3}} = \Delta x_1 \gamma_0^{\frac{1}{3}} , \qquad \overline{\Delta x_1'} = V_1'^{\frac{1}{3}} = V_2^{\frac{1}{3}} \gamma_0^{\frac{1}{3}} = \Delta x_2 \gamma_0^{\frac{1}{3}} . \tag{20}$$

Imagine two real material bodies ($m \neq 0$) being spherically symmetrical and identical in all aspects, their centers resting in the coordinate sources of $S_1$ and $S_2'$, to collide at $\Sigma_0$ at the time $t_2' = t_1 = t_0 = 0$. At this moment (7) and (16) are fully valid with the consequence that the x-dimension of the body in $S_2'$ appears still altered relativistically - as observed from $S_1$ - so that its mean effective scattering volume at the time of collision must be

$$\overline{V_2'} = \overline{\Delta x_2'} \Delta y_2' \Delta z_2' = \overline{\Delta x_2'} \Delta y_1 \Delta z_1 \tag{21}$$

and in the inverse case:

$$\overline{V_1'} = \overline{\Delta x_1'} \Delta y_1' \Delta z_1' = \overline{\Delta x_1'} \Delta y_2 \Delta z_2 . \tag{22}$$

In connection with (20) we receive

$$\overline{V_2'} = V_1 \gamma_0^{\frac{1}{3}} , \qquad \overline{V_1'} = V_2 \gamma_0^{\frac{1}{3}} . \tag{23}$$

Thus, as deduced from either system, considered to be at rest, in ultra relativistic collision events the body resting in the system considered moving must in the mean seem enlarged by the factor $\gamma_0^{1/3}$. Hence its mean relative geometrical interaction-radius averaged over the three spatial dimensions must be
wherefrom the mean geometrical cross-sections follow:

$$\overline{\sigma}'_2 = \pi(\Delta r_1 \gamma_0^{\frac{1}{9}})^2 = \sigma_1 \gamma_0^{\frac{2}{9}}, \qquad \overline{\sigma}'_1 = \pi(\Delta r_2 \gamma_0^{\frac{1}{9}})^2 = \sigma_2 \gamma_0^{\frac{2}{9}}. \tag{25}$$

$$\overline{\Delta r'_2} = V_1^{\frac{1}{3}} \gamma_0^{\frac{1}{9}} = \Delta r_1 \gamma_0^{\frac{1}{9}}, \qquad \overline{\Delta r'_1} = V_2^{\frac{1}{3}} \gamma_0^{\frac{1}{9}} = \Delta r_2 \gamma_0^{\frac{1}{9}}, \tag{24}$$

In either system the colliding bodies rest, the same enhancement of the interaction-radius (24) or of the geometrical cross-section (25) with growing velocity will be noticed. Therefore, relative to the kinematic center $\Sigma_0$, which at at the time $t'_2 = t_1 = t_0 = 0$ coincides with $S_1$ and $S_2$, the mean total geometrical cross-section is given by

$$\overline{\sigma}_{geo} = \overline{\sigma}'_2 + \overline{\sigma}'_1 = 2\overline{\sigma}_1 \gamma_0^{\frac{2}{9}}, \tag{26}$$

whereby from the foregoing it is clear that $\overline{\sigma}'_2 = \overline{\sigma}'_1$. As is shown below, (24) and (26) are directly related to the interaction radii as derived from high energetic collisions on the strength of the optical theorem and the total cross-sections, respectively.

Consider a particle, based at a system $S_2^\circ$, moving through a dense material medium at ultra relativistic velocity and coming to a halt within the medium. Relative to $S_2^\circ$ the atoms, constituting the medium and resting relative to another, obviously represent the "real resting" system $S_1$ according to (7). According to (18) and (19) a moving particle $S_2^\circ$ would relative to the resting atoms seem enhanced by the factor $\gamma_0^2$ - measurements by light signals. Analogous application of (20) to (25) leads to the geometrical cross-section

$$\overline{\sigma}_2^\circ = \sigma_1 \gamma_0^{\frac{4}{9}}, \qquad \overline{\sigma}_1^\circ = \sigma_2 \gamma_0^{\frac{4}{9}}. \tag{27}$$

This result predicts a shrinkage of the interaction mean-free-paths of particles plunging through some material before coming to a halt:

$$\frac{\overline{\sigma}_2^\circ}{\sigma_1} = \frac{\lambda_1}{\lambda_2^\circ}, \qquad \lambda_2^\circ = \lambda_1 \gamma_0^{-\frac{4}{9}}, \tag{28}$$

where $\lambda_1$ means mean-free-path of a slowly moving particle and $\sigma_1$ the mean geometrical cross-section if $v_0 \ll c$.

## 6. Experimental Interaction-Radii, Total Cross-Sections and Mean-Free-Paths Compared with Theory

In the following is investigated, whether the experimentally found rise of the interaction-radius of the protron on the strength of the optical theorem is in accord with (24) and, furthermore, the total cross-section $\sigma_{tot} = \sigma_{el} + \sigma_{inel}$ of hadrons at ultra relativistic collisions in colliders possibly depends solely on the total geometrical cross-section according to (26). We restrict to protrons and antiprotrons, which we assume to be (only) geometrically alike.

The geometrical cross-section of the "resting" or "slowly moved" protron (antiprotron) is measured with $\approx 1(\text{fm})^2 = 10^{-26}\,\text{cm}^2 = 10\,\text{mb}$ (millibarn). Consequently, according to (24) the mean interaction-radius of the protron (antiprotron) amounts to (in Fermis)

$$\overline{\Delta r_{geo}} = 1 \times \gamma_0^{\frac{1}{9}} = (1 - \beta_0^2)^{-\frac{1}{18}} \tag{29}$$

and the mean total geometrical cross-section according to (26) rises to (in millibarn)

$$\overline{\sigma_{geo}} = 2 \times 10 \times \gamma_0^{\frac{2}{9}} = 20 \times (1 - \beta_0^2)^{-\frac{1}{9}}, \tag{30}$$

irrespective of quantum-mechanical effects. Effects of spin are considered to average out over a wide range of collision events.

But before computing the geometrical cross-sections and interaction-radii, we have yet to conceive fair approximations of $N_i$ and, therewith, of $\gamma_0$ as a function of the center-of-mass energy to render those computations .
According to (14) the relativistic momentum is given by

$$p_0 c = \sqrt{E'^2 - E^2} \approx E'$$

if $E' \gg E$, where E means rest energy. If $\beta_0 \to 1$ in connection with (6) is deduced

$$\beta_0 \gamma_0 \approx \gamma_0 \approx \frac{\beta \gamma}{N_i} \approx \frac{\gamma}{N_i} \approx \frac{E'}{mc^2} = \frac{E^*}{mc^2 N_i},$$

where E* means total center-of-mass energy. Thus, we can write

$$\gamma_0^{\frac{1}{9}} \approx \left(\frac{E'}{mc^2}\right)^{\frac{1}{9}} \approx \left(\frac{E^*}{mc^2 N_i}\right)^{\frac{1}{9}} \text{ as well as } \gamma_0^{\frac{2}{9}} \approx \left(\frac{E^*}{mc^2 N_i}\right)^{\frac{2}{9}}.$$

From the relativistic addition theorem follows

$$\gamma_0 \approx 2\gamma_1^2 \approx 2^3 \gamma_2^4 \approx 2^7 \gamma_3^8 \text{ etc.},$$

when $\beta_0 \to \beta_1 \to \beta_2 \to \beta_3 \to 1$. According to (6a) in the case of $N_1 - N_3$ is valid $N_1^{2/9} \to N_1^{1/9} \to 1$. Therefore, to compute $N_1^{1/9} - N_3^{1/9}$, or $N_1^{2/9} - N_3^{2/9}$, respectively, it will be only a minor error to deduce $\beta_1 - \beta_3$ from the approximations

$$2\gamma_1^2 \approx \frac{E_1^*}{mc^2}, \quad 2^3\gamma_2^4 \approx \frac{E_2^*}{mc^2}, \quad 2^7\gamma_3^8 \approx \frac{E_3^*}{mc^2}$$

if the respective $E^*$ successively is reached, where $E_3^* > E_2^* > E_1^*$. Therefrom is given:

$$\beta_1^2 \approx 1 - \frac{2mc^2}{E_1^*}, \quad \beta_2^2 \approx 1 - \left(\frac{8mc^2}{E_2^*}\right)^{\frac{1}{2}}, \quad \beta_3^2 \approx 1 - \left(\frac{128mc^2}{E_3^*}\right)^{\frac{1}{4}}. \qquad (31)$$

Inserting the respective result of (31) into (6a) yields $N_1 - N_3$. If $E^*$ is given in GeV, for protons (antiprotons) in the case of colliders follows

$$\gamma_0 \approx E' \approx 2 \times \frac{E^*}{2 \times N_i} \qquad (32)$$

and in the case of accelerators

$$\gamma_0 \approx E' \approx 2 \times \frac{E^*}{(1+\beta_1^2) \times N_i}, \qquad (33)$$

where $1 + \beta_1^2$ in (33) is due to the relativistic addition of $2E^*$, implying $N_i$ to start with $1 + \beta_2^2$. Inserting $N_i$ - computed from (6a) - into (32), or (33), yields $\gamma_0$ as a function of $E^*$ and, therewith, the interaction-radius according to (29) or the total geometrical cross-section according to (30). In the following some exemplary theoretical computations are compared with experiment.

TABLE I: INTERACTION-RADII (accelerators, last line: collider)

| $E^*$ (GeV) | $N_i$ | $E' \approx \gamma_0$ (GeV) | $\Delta \bar{r}_{geo} = \gamma_0^{1/9}$ (fm) | $\Delta r$ experiment (fm) [5] |
|---|---|---|---|---|
| 5   pp | $N_1 \approx 1.60$ | 6.25 | 1.23 | $\approx 1.23$ |
| 6   pp | $N_1 \approx 1.67$ | 7.18 | 1.25 | $\approx 1.25$ |
| 8   pp | $N_1 \approx 1.75$ | 9.14 | 1.28 | $\approx 1.28$ |
| 10  pp | $N_1 \approx 1.80$ | 11.11 | 1.31 | $\approx 1.31$ |
| 20  pp | $N_1 \approx 1.90 \times 1.37$ | 15.39 | 1.35 | $\approx 1.36$ |
| 30  pp | $N_1 \approx 1.93 \times 1.48$ | 20.95 | 1.40 | $\approx 1.40$ |
| 50  pp | $N_1 \approx 1.92$ | 26.04 | 1.44 | $\approx 1.43$ |

TABLE II: CROSS-SECTIONS (colliders)

| $E^*$ (GeV) | $N_i$ | $E' \approx \gamma_0$ (GeV) | $\bar{\sigma}_{geo} = 20 \times \gamma_0^{2/9}$ (mb) | $\sigma_{tot}$ experiment (mb) |
|---|---|---|---|---|
| 62    pp$^-$ | $N_1 \approx 1.94$ | 32.03 | 43.21 | $43.9 \pm 0.6$ [6] |
| 546   pp$^-$ | $N_2 \approx 3.64$ | 149.93 | 60.89 | $61.9 \pm 1.5$ [7,8] |
| 900   pp$^-$ | $N_3 \approx 4.16$ | 216.58 | 66.08 | $\approx 66$ [9] |
| 1800  pp$^-$ | $N_3 \approx 4.97$ | 362.19 | 74.08 | $78.3 \pm 5.9$ [10,11] |
| 10000 | $N_3 \approx 6.20$ | 1613.74 | 103.25 | ----- |
| 14000 | $N_3 \approx 6.36$ | 2200.83 | 110.62 | ----- |
| 40000 | $N_3 \approx 6.77$ | 5904.22 | 137.74 | ----- |

Table II includes predictions for CERN's Large Hadron Collider ($E^* = 10$ TeV in 2005 and 14 TeV in 2008) and for the cancelled Superconducting Super Collider, which could have reached 40 TeV. It seems that from the kinematical region $E^* \approx 60$ GeV on the total cross-sections of protrons and antiprotons coincide, to depend solely on their mean geometrical cross-sections. In the case of protrons as a component of the cosmic radiation equ. (30) in connection with (33) predicts at $E^* = 44$ TeV a mean geometrical cross-section of $\approx 160$ mb, in good accord with experiment [12].
It is predicted that the enhancement of the geometrical cross-section or interaction-radius also delivers an explanation of the "EMC-effect".

According to (28) the interaction mean-free-paths of nucleii or particles coming to a stop within some material (e.g. nuclear track emulsion), after traversing it at ultra relativistic velocity (energy), must shrink proportional to the factor of $\gamma_0^{-4/9}$.
The EMU 08 experiment at CERN studied the interactions of oxygen beams at E* = 200 and 60 GeV/nucleon in nuclear emulsion and found for inelastic events the interaction mean-free-paths for higher and lower energy beams to be 10.89 cm and 12.84 cm [13]. Extrapolation according to (28) results in 10.89 cm, too.
For secondary particles this effect is experimentally well-known and controversially discussed under the term "anomalons" (secondary nuclei with abnormal short mean-free-paths after collision of primaries within some material).

## 7. Further Physical Implications

Suppose a body of mass m' = **V**'ρ', resting in the coordinate source of the frame of reference S' - where **V**' means volume as defined by (16) and ρ' density of mass -, to move inertially relative to the frame of reference S, considered to be at rest. From (11) follows m' = m$\gamma_0$ and, therewith,

$$\mathbf{V}'\rho' = \mathbf{V}\rho\gamma_0. \qquad (34)$$

Inserting (16) into the left hand member of (34) yields

$$\mathbf{V}\rho'\gamma_0 = \mathbf{V}\rho\gamma_0, \qquad (35)$$

implying ρ' = ρ = const. Thus, it must be valid

$$E'^2 - E^2 = E'^2\beta_0^2 = \mathbf{V}'^2\rho^2 c^2 v_0^2,$$
$$E'\beta_0 = \mathbf{V}'c v_0 \times (\rho = \text{const}),$$
$$= d\mathbf{t}' c v_0 \times (\rho \times c \times dydz = \text{const}),$$

where **V**' = dx'dy'dz' = dx' × dydz. Because ρ' = ρ = const, dx' × dydz = dt' × c × dydz and c × dydz = const, too, the relativistic growing of E'$\beta_0$ = m$\gamma_0$c$v_0$ must be caused by the relativistically dilated time d**t**' = dt$\gamma_0$ alone. If we write

$$d\mathbf{t}' c v_0 = E_t = m_t c^2,$$
$$m_t = \frac{E_t}{c^2} = \frac{d\mathbf{t}' v_0}{c} = \frac{d\mathbf{x}'}{c}, \qquad (36)$$

where "$E_t$" means energy of dilated time dt'c$v_0$ of a moving material body and "$m_t$" mass induced by time dilation, we see clearly that the fraction $v_0/c$ of the dilated time is the relevant factor, which produces the relativistic mass of relativistic kinematics and vanishes if $v_0 \to 0$.

Considering the definition of action as the product of energy and time, we write

$$E \times \tau_0 = mc^2 \times \tau_0 = h, \tag{37}$$

where "$\tau_0$" denotes quantum of time according to (5) and "h" Planck's constant, the quantum of action. Equ. (37) delivers

$$\lambda_0 \times mc = h, \tag{38}$$

where $\lambda_0 = \tau_0 c$ means fundamental length or minimum distance. Because $\lambda_0$ and h are minima, m must be a minimum, too. From (36) we receive $m_t c = d\mathbf{x}'$ so that in the case of $mc = m_t c$ (38) attains the form

$$\lambda_0 \times d\mathbf{x}' = h$$

and because $mc = d\mathbf{x}'$ = Min. it follows $d\mathbf{x}' = \lambda_0$ and, thus,

$$\lambda_0 \times mc = \lambda_0^2 = \tau_0^2 c^2 = m^2 c^2 = h. \tag{39}$$

Hence the numerical value of the minimum distance must be

$$\lambda_0 = \sqrt{h} = 0.814013 \times 10^{-13} \text{ cm} \tag{40}$$

(state of rest in the rest frame) and of the quantum of time

$$\tau_0 = \frac{\sqrt{h}}{c} = 2.715256 \times 10^{-24} \text{ s}, \tag{41}$$

whereby $h = 6.626176 \times 10^{-27}$ erg × s and $c = 2.99792458 \times 10^{10}$ cm × s$^{-1}$. This result is in accord with the CGS-system, where h is defined by erg × s = 1g × (cm)$^2$/s$^2$ × s. According to (39) is valid m (g) = $\lambda_0/c$ = $\tau_0$ (s) so that $\sqrt{h}$ really results in $\lambda_0$ cm. The existence of a fundamental unit of length of the order of magnitude 10$^{-13}$ cm, which has been first proposed by Werner Heisenberg nearly 60 years ago [14], also explains Heisenberg's uncertainty principle: an uncertainty smaller than $\Delta E \times \Delta t = h = \lambda_0^2$ cannot exist. Furthermore, it is clear that Lorentz transformations with coordinate differences smaller than $\lambda_0$ and $\tau_0$ are not possible.

If the mean life-times of short-lived elementary particle resonances are divided by $\tau_0$

$$\frac{\overline{T}}{\tau_0} = \frac{h}{\tau_0 \Gamma} = n, \; (n = 1, 2...),$$

where T means life time and Γ full width, in by far the most cases nearly integers and in the others integers plus a half result, e. g. 0.98 for the top quark (computed Γ ≈ 1.55 GeV) implying its life-time be exactly one quantum of time, and 3.95 for the 1370 MeV "exotic" meson (Γ ≈ 385 MeV), recently found at Brookhaven's AGS [15].

Let $\lambda_0$ be smallest part of an one-dimensional manifold $R^1$. Then, necessarily, the smallest possible triangle in $R^2$ is a Pythagorean one with the small sides $1 \times \lambda_0$ and the hypotenuse $(1 \times \lambda_0^2 + 1 \times \lambda_0^2)^{1/2} = \sqrt{2} \times \lambda_0$, being the fundamental length in $R^2$. Consequently the fundamental length in $R^3$ must be $(1 \times \lambda_0^2 + 2 \times \lambda_0^2)^{1/2} = \sqrt{3} \times \lambda_0$. Therefore, the classical electron radius $r_e$ in $R^3$ must be a symmetric multiple of $\sqrt{3} \times \lambda_0$. If we put $r_e = 2 \times \sqrt{3} \times \lambda_0$ this results in $2.8198 \times 10^{-13}$ cm as compared with $r_e = e^2/(4\pi\varepsilon_0 \times m_e c^2) = 2.8179 \times 10^{-13}$ cm.

Apart from a numerical factor, can $m = \lambda_0/c = \tau_0 =$ Min. only be interpreted to express the rest mass of the smallest piece of stable and electrically neutral matter, the hydrogen atom in its ground state. Furthermore, leads the similarity between $m_t = E_t/c^2 = d\mathbf{x'}/c$ and $m = E/c^2 = \lambda_0/c$ to the conclusion that relativistic mass and rest mass of electrically neutral matter must be of the same origin.

It is clear that from the foregoing result further implications yet, especially for high energy physics, but which could not dealt with in this first sketchy concept of naturally composite velocity. I thank my wife Ingrid as well as my friend Dr Paul Yule for their assistance and Benjamin Kunst for helpful discussions, which contributed much to clarify the basic idea of this study.

## References

 [1] Einstein, A., Ann. d. Phys. **17**, 906, (1905)
 [2] Einstein, A., Ann. d. Phys. **17**, 895 - 897, (1905)
 [3] Einstein, A., Gundzüge d. Relativitätstheorie, 5th ed., Vieweg-Verlag, Braunschweig 1969, p. 31
 [4] Laue, M. v., Ann. d. Phys. **23**, 989, (1907)
 [5] Amaldi, U., Scientific American, 43, (November 1973)
 [6] Breakstone, A., et al., Nucl. Phys. B **248**, 253, (1984)
 [7] Bernard, D. et al., Phys. Lett. B **198**, 583, (1987)
 [8] Alner G. J. et al., Phys. Lett. **138** B, 304, (1984)
 [9] CERN Courier, 14, (May 1990)
[10] Amos, N. A. et al., Phys. Lett. B **247**, 127, (1990)
[11] Amos, N. A. et al., Phys. Rev. Lett. **63**, 2784, (1989)
[12] CERN Courier, 58, (March 1984)
[13] CERN, Annual Report **II**, 32, (1987)
[14] Heisenberg, W., Ann. d. Phys. **32**, 20 - 32, (1938)
[15] Thompson, D. R. et al., Phys. Rev. Lett. **79**, 1630 - 1633, (1997)